\documentclass[pra,twocolumn,aps,amssymb,footinbib]{revtex4}
\usepackage{amssymb}
\usepackage{graphicx}
\usepackage{amsmath}
\usepackage{times}
\usepackage{color}
\usepackage{subfigure}
\usepackage{setspace}
\usepackage{bm}

\begin{document}

\begin{abstract}
We propose a controlled method to create and detect d-wave superfluidity
with ultracold fermionic atoms  loaded in  two-dimensional optical
superlattices. Our scheme consists in preparing an array of
nearest-neighbor coupled square plaquettes or ``superplaquettes'' and using  them as  building blocks to
construct a d-wave superfluid state. We describe  how to use  the coherent
dynamical evolution in such a system to experimentally probe the pairing
mechanism. We also derive the zero temperature phase diagram of the
fermions in a  checkerboard lattice (many weakly coupled plaquettes)
and show that by tuning the inter-plaquette tunneling spin-dependently
or varying the filling factor one can drive the
 system into a d-wave superfluid phase or a Cooper pair density wave phase.
We discuss  the use of noise correlation measurements to experimentally
probe these phases.

\end{abstract}

\title{Preparation and detection of  d-wave superfluidity in two-dimensional  optical
superlattices}

\author{ A. M. Rey$^{1}$,  R. Sensarma$^{2}$,  S. Foelling$^{2}$, M. Greiner $^{2}$, E. Demler$^{1,2}$ and M.D.
Lukin$^{1,2}$}  \affiliation{$^{1}$ Institute for Theoretical Atomic, Molecular and Optical Physics,
Harvard-Smithsonian Center of Astrophysics, Cambridge, MA, 02138.} \affiliation{$^{2}$ Physics Department,
Harvard University, Cambridge, Massachusetts 02138, USA}
\maketitle
\section*{}
Ultracold atoms in optical lattices\cite{Opt_Lat} are promising
simulators of complex many-body problems and model Hamiltonians that arise
in condensed matter physics. They provide a clean system with parameters
which can be tuned in a controlled fashion from the weakly interacting to the
strongly interacting limits. The observation of the superfluid to Mott
insulator transition with bosons\cite{Greiner_1} and recent experimental
realization of both repulsive and attractive Hubbard
models\cite{Essl_1,Essl_2} with Fermions are important steps in that
direction. Recently, the existence of superexchange
antiferromagnetic correlations have been shown in an array of isolated
double wells\cite{rey,Bloch}. Here we extend similar considerations for a d-wave superfluid
state and propose an experimental scheme to realize and detect it.

The repulsive Hubbard model on a square lattice is one of the most important
model Hamiltonians for strongly interacting fermions and is widely believed
to contain the essential features of high temperature
superconductors\cite{And_Sc}. At half filling (one fermion per site),
strong repulsion leads to localization of the fermions
resulting in a Mott insulator. Superexchange interactions lead to an
antiferromagnetic ground state. Away from half filling, antiferromagnetic
correlations compete with the kinetic energy of the holes and the phase
diagram remains unknown. Although there is no rigorous proof of existence
of superfluidity in this model, approximate theoretical and  Monte Carlo
simulations at high temperatures \cite{Bickers}  have conjectured the
presence of d-wave superfluid ground states away from half-filling.

\begin{figure}[h]
\begin{center}
\leavevmode {\includegraphics[width=3in]{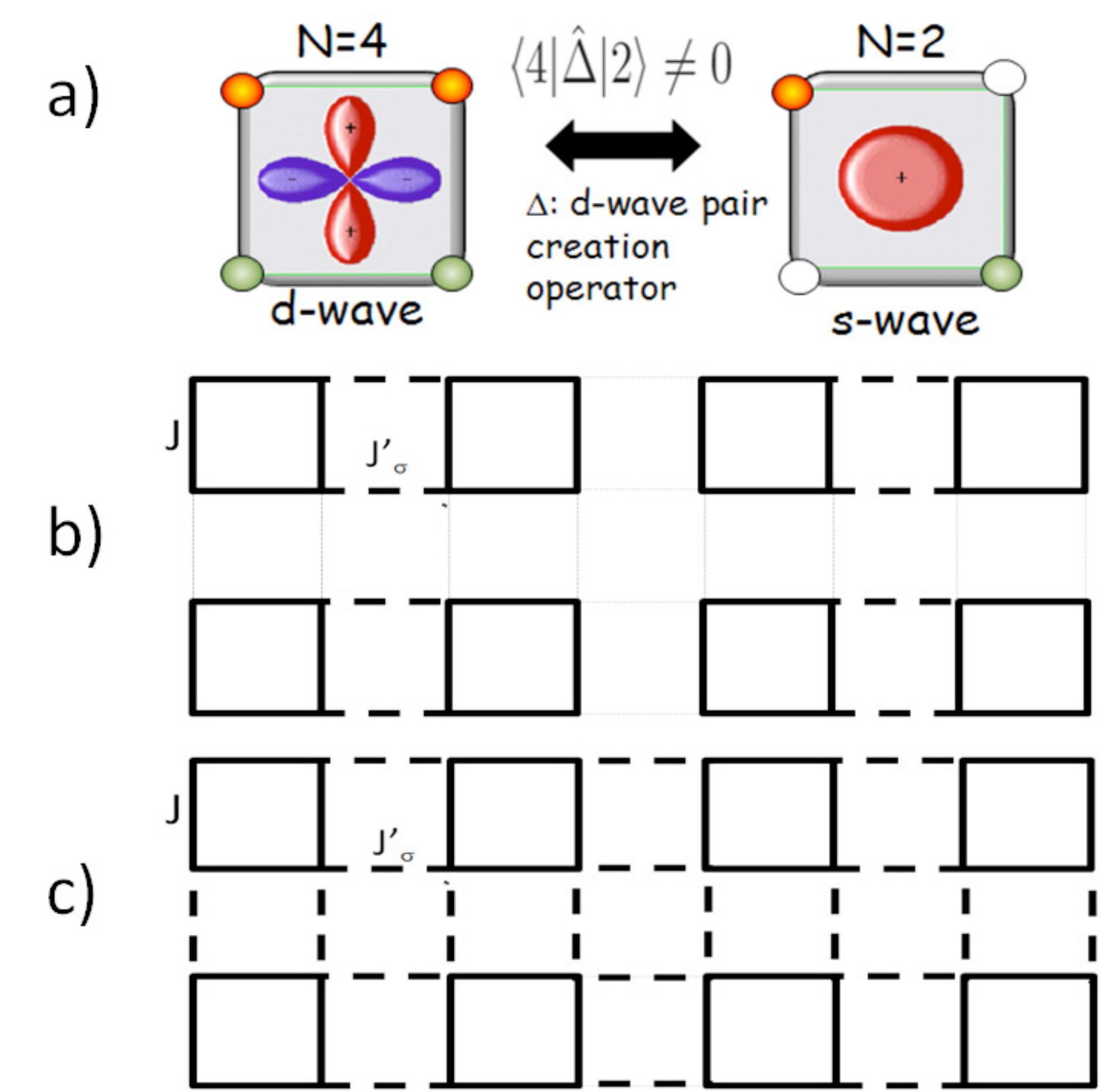}}
\end{center}
\caption{ A plaquette is the minimum system that exhibits d-wave symmetry. a) When loaded with four fermions the ground state is d wave symmetric while when loaded with 2 the ground state exhibits s wave symmetry. Consequently the two states have non zero matrix element with the d-wave pair creation operator.  Here we consider the situations when  plaquettes are coupled into a  super-plaquette array and  into   a  checkerboard array, which are schematically represented in b) and c) respectively.
In the picture the spin independent  intra-plaquette tunneling  $J$ is represented by  a thick solid  line and the inter-plaquette tunneling, $J'_\sigma\ll J$  by   a dashed line. The subscript  $\sigma$ in  $J'_\sigma$ emphasizes that it can  depend on the spin of the atom. }\label{superp}
\end{figure}

It has been known for some time\cite{Scalapino} that the minimal 2D unit
that can sustain d-wave pairing physics is a single square plaquette
(see Fig.\ref{superp}).
The single plaquette has been the starting point of various
approaches\cite{CORE,Altman2,Kivelson,Trebst} to the Hubbard model on the
square lattice. Even with purely repulsive
interactions, there is a range of parameters where two holes tend to bind
together on a single plaquette rather than to delocalize among  different
plaquettes. The hole pair that is created has a d-wave symmetry and can
lead to d-wave superfluidity once the plaquettes are coupled.

\begin{figure*}
\centering
{\includegraphics[width=6in]{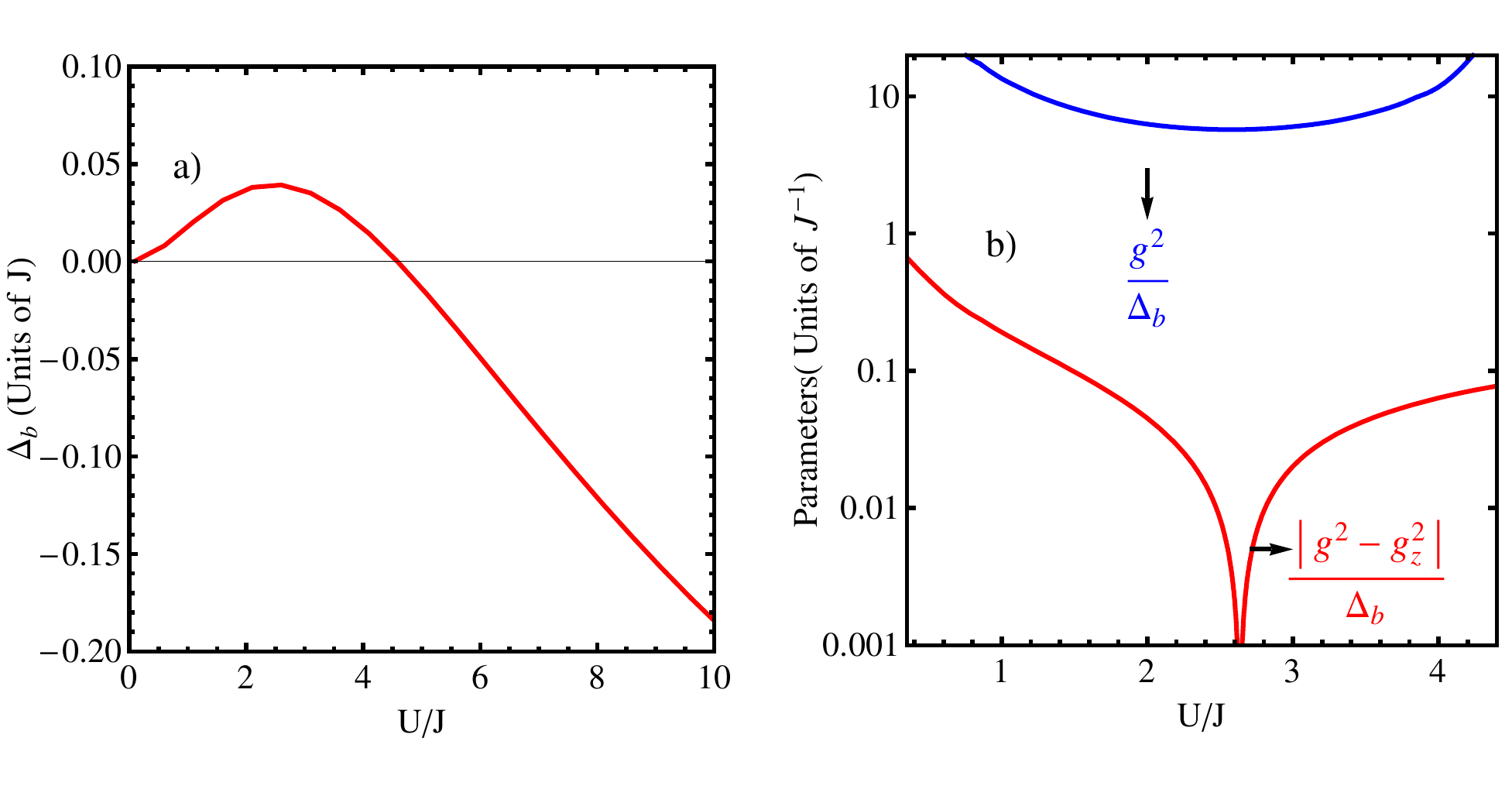}}\quad
\caption{  a)  The red solid line corresponds to the pair binding energy in a plaquette. For $0<U/J<4.6$, $\Delta_b>0$ and consequently it is energetically favorable to have two holes in the same plaquette.  b)  Coupling parameters as function of $U/J$ for the  effective XXZ Hamiltonian, Eq. (\ref{xxz}). The blue line corresponds to $g^2/\Delta_b$ as a function of $U/J$. This is the only parameter that appears in the super-plaquette Hamiltonian. When more than two plaquettes are coupled, new virtual processes have to be accounted for which  change the  Ising term coupling constant. While the latter become  proportional to  $g_z^2/\Delta_b$,  the transverse  coupling constant remains the same, i.e. proportional to  $g^2/\Delta_b$. The red line shows the absolute value of the difference between these two  parameters.} \label{bindi}
\end{figure*}

We first propose an experimental scheme to verify
these concepts by creating an array of isolated super-plaquettes (i.e. two
adjacent plaquettes coupled by a weak tunneling, see Fig. \ref{superp}b) and
loading them with six Fermions each. To measure the binding energy of the hole
pair, we analyze the coherent dynamics of the super-plaquette.
Finally we derive the phase diagram of the weakly coupled plaquettes with
spin-dependent inter-plaquette tunneling. We show that
the system can be driven to either a d-wave superfluid or a density wave
state\cite{Kivelson}. We propose an experimental scheme to realize this and
discuss the detection  of the two  quantum phases via atomic noise
correlations measurements\cite{Altman,Foelling,Greiner,Rom,Spielman}.

\section{Plaquette fermion models}

Consider fermions in an isolated plaquette (shown in Fig.\ref{superp}a).
Assuming one accessible single particle state in each well (i.e. level spacing much larger
than other energy scales in the problem), the system is described by the
Hubbard Hamiltonian
\begin{equation}
\hat{H}= - J \sum _{\langle r,r'\rangle,\sigma}\hat{c}_{ r\sigma }^{\dagger}\hat{c}^{}_{r'\sigma}+U \sum_{r}
 \hat{n}_{\uparrow r}\hat{n}_{\downarrow,r} \label{plaq}
\end{equation}
where $J$ is the tunneling matrix element and $U$ is the onsite Hubbard
repulsion. Here $\hat{c}^{}_{r\sigma}$ are fermionic annihilation operators,
$\hat{n}_{r\sigma }=\hat{c}^{\dagger}_{r\sigma } \hat{c}^{}_{r\sigma }$ are
number operators, $r=1, \dots 4$, labels the four  sites in a plaquette
and the term  $\langle r,r'\rangle$  indicates that the sum is restricted
to nearest neighbors.

 The eigenstates in a single plaquette depend on the filling factor.
When  filled with   $N=4$ or $N=2$ fermions, the ground state is a
spin-singlet exhibiting $d$ and $s$ wave symmetry respectively. We denote
the corresponding ground states as $|4\rangle$ and $|2\rangle$
(See Fig.\ref{superp}a).
On the other hand, for $N=3$, the ground state is degenerate with $S=1/2$
and $p_x\pm i p_y$ symmetry in the regime $U<U_t\sim 18.6 J$. We denote
them  as $|3^{(\sigma,\tau)}\rangle$ with $\sigma$ and
$\tau$ specifying the spin polarization and the orbital "chirality"
($\tau=\pm$). The $d$ vs $s$ wave symmetry of the $|4\rangle$ and
$|2\rangle$ states is the crucial element in obtaining the d-wave
pairing mechanism, and the hole-pair creation operator that connects the two
states must have a d-wave symmetry\cite{Scalapino}; i.e.
$\langle 2|\hat{\Delta}^\dagger_d|4\rangle \neq 0$, where
\begin{equation}
\hat{\Delta}_d=(\hat{s}_{12}+\hat{s}_{34}-\hat{s}_{14}-\hat{s}_{23})/2, \label{dwave}
\end{equation}
and $\hat{s}_{rr'}=(\hat{c}^{\dagger}_{r\uparrow}\hat{c}^{\dagger}_{r'\downarrow }
-\hat{c}^{\dagger}_{r\downarrow}\hat{c}^{ \dagger}_{r'\uparrow})/\sqrt{2}$
creates a singlet on the $rr'$ bond.

 \begin{figure*}
\centering
{\includegraphics[width=6 in]{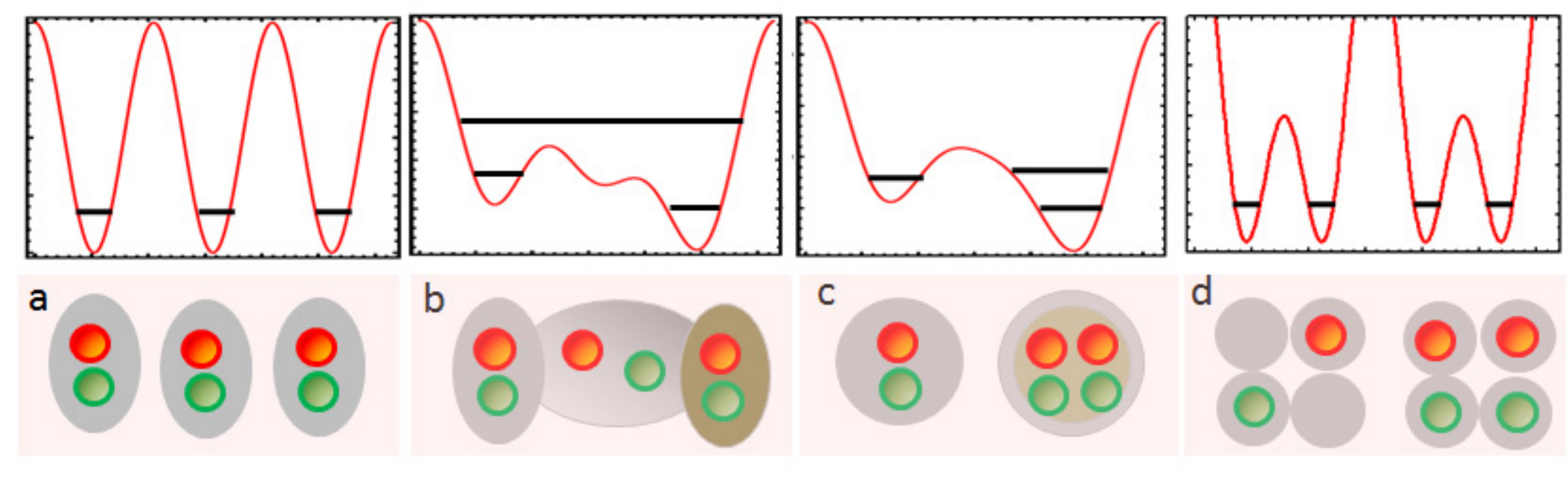}}
\centering
\caption{Loading of fermions in a patterned loading array of plaquettes:
a) First  a unit filled band insulator  of fermionic atoms  is prepared  in a  rectangular lattice with lattice period $4/3 \lambda$ and $2\lambda$ along the $x$ and $y$ direction respectively.
b)  A  phase shifted $4 \lambda$  lattice along  the x axis is slowly  ramped up.
c)As the  period $2\lambda$ lattice is ramped up, the period $4/3 \lambda$ is ramped down generating a double well lattice. The bias is  chosen such that the  ground state  has   four fermions in the right and two fermions in the left  sites of each double well.
d) The lattice depth of the  $2\lambda$ lattice is increased to make  all wells independent. Subsequently   the bias  is   turned off  and  the  isolated wells  split into two. The overall scheme creates an array of independent plaquettes loaded in the ground state with alternating filling factors of $2$ and $4$ along the x direction.}
\label{load}
\end{figure*}

We next look at the case of $2$ holes in two isolated plaquettes. The holes
can bind together within the same plaquette or separate as single holes in
each plaquette depending on the binding energy defined as
\begin{equation}
\Delta_b=2 E_g(N=3)-E_g(N=4)- E_g(N=2)
\end{equation}
being positive or negative respectively. Here $E_g(N=n)$ is the single
plaquette ground state energy when loaded with $n$ atoms.
As shown in Fig. \ref{bindi}, $\Delta_b$ is a non-monotonic
function of $U/J$ \cite{Altman2,Kivelson,Trebst,Schumann}, which
reaches a maximum value of $\Delta_b\approx 0.04 J$  at $U\approx2.45J$
and becomes negative for $U=U_c> 4.58 J$. Consequently only when $U<U_c$
hole pairs on a single plaquette are energetically stable.

Two adjacent plaquettes can be coupled through a weak (possibly spin
dependent) tunneling $J'_\sigma$ to form a super-plaquette
(see Fig.\ref{superp}). As long as $0<J'_\sigma\ll \Delta_b$, the states
$|4,2\rangle$ and  $|2,4\rangle$ are lower in energy and occupation of the
states $|3^{(\sigma,\tau)},3^{(\bar{\sigma},\bar{\tau})}\rangle$ are energetically
suppressed. They can only be populated
as "virtual" intermediate states leading to an effective super-exchange
interaction between $|4,2\rangle$ and  $|2,4\rangle$. Specifically, by
treating the $|4\rangle$ and $ |2\rangle$ states as  the
pseudo-spin components $|\Uparrow\rangle$ and $|\Downarrow\rangle$ of an
effective  spin $1/2$ system, the interaction between the effective spins
can  be described by an  XXZ-type  Hamiltonian
\begin{equation}
H_{eff}=-\frac{ J_\uparrow'J_\downarrow' g^2}{\Delta_b}(\sigma_R^x\sigma_L^x+\sigma_R^y\sigma_L^y)+ \frac{g^2(
J_\uparrow'^2+ J_\downarrow'^2 ) }{2\Delta_b}\sigma_R^z\sigma_L^z \label{tpl}
\end{equation}
with $\sigma_{i=R,L}^{\alpha=x,y,z}$ standard Pauli matrices acting on the
right (R) or left (L) effective pseudo-spins and
$g$ the coupling matrix element between the right and left
plaquettes which is of order one (see Fig.\ref{bindi}).

Eq.(\ref{tpl}) contains the essential physics  we are interested in this work.
If $ J_\uparrow'=J_\downarrow'$ the energy eigenstates are effective triplet
and singlet states $|t,s\rangle=\frac{1}{\sqrt{2}}(|4,2\rangle\pm|2,4\rangle)$
with the triplet being  the ground state. These are separated by an energy
gap $\sim\Delta_b$ from the rest of the Hilbert space. The ground state
has a  non zero expectation value of the {\it d-wave} pair correlation operator
$\langle t|\hat{\Delta}^\dagger_R \hat{\Delta}_L|t\rangle\neq 0$ which leads to a
d-wave superfluid when coupling all the plaquettes. On the contrary if
$ J_\uparrow'\ll J_\downarrow'$, the Ising term dominates and any infinitesimal
symmetry-breaking perturbation will collapse the state into
$|4,2\rangle$ or $|2,4\rangle$ which are inherently density-ordered states.
 These considerations indicate that the many body phase diagram of this model
would depend on the spin-dependent couplings in a non-trivial way.

\section{Preparation and Detection }

To verify the energy structure of the hole-pair states
we want to create an array of isolated super-plaquettes loaded with $2$ and $4$
fermions in the left and the right plaquettes respectively.

An array of plaquettes can be created by superimposing two orthogonal optical super-lattices formed by the superposition of two independent sinusoidal potentials which differ in periodicity by a factor of two \cite{Bloch,Sebby,Paredes}. The aim here is first to load  the fermions in a 2D array of independent
plaquettes with alternating filling factors of 4 and 2 along one direction.
The preparation procedure we propose is based on a unity-filled band
insulator \cite{Kohl}, and uses adiabatic manipulations of a super-lattice
potential created by standing waves with four different periodicities
$\lambda$, $2\lambda$, $4\lambda$ and  $4\lambda/3$ (For other strategies
see \cite{Trebst}). Such wavelength combinations are available for
typical Fermionic atoms such as ${}^{40}$K and ${}^6$Li or can be
engineered by intersecting four pairs of laser beams with appropriate
angles \cite{Peil}, resulting in a set of four equidistant $k$-vectors. For
the discussion we will also assume that there is a deep  axial lattice
that freezes the atom motion along the $z$  direction.

The process of patterned loading an array of plaquettes is depicted in
Fig.\ref{load}. The  initial band insulator is formed in a  rectangular
lattice with period $4/3 \lambda$ and $2\lambda$ along the $x$ and $y$
directions, respectively. Adiabatically introducing a  phase shifted
$4\lambda$ lattice along the $x$ axis creates an effective
4$\lambda$-lattice with three sites per cell with different energy
offsets (See Fig.\ref{load}). By ramping down the period $4/3 \lambda$
lattice while ramping up the period $2\lambda$ one can convert the potential
into a double well lattice with an asymmetric energy offset. This offset
 has to be large compared to the on-site repulsion to guarantee that
the final ground state corresponds to a system with four fermions in the
lower and two fermions in the higher sites of each double well. After
increasing  the lattice depth of the  $2\lambda$ lattice to suppress
tunneling between the wells the bias can be removed and the isolated wells
can be split into four sites by slowly turning on the $ \lambda$ lattice
along both x and y directions. As a result, an array of independent
plaquettes in the ground state with alternating filling factors of
$4$ and $2$ along the $x$ direction is created. To obtain an array of
super-plaquettes, the tunneling between adjacent
plaquettes can be controlled by the depth of the $2\lambda$ potential, while
the $4\lambda$ lattice isolates super-plaquettes. The extra $4\lambda/3$ lattice is needed to balance the offset created when the long lattice is added. Spin dependent control of the inter-plaquette tunneling can be achieved
by additional control over the laser polarizations \cite{Sebby}.

\begin{figure}
\centering
\centering
{\includegraphics[width=3in]{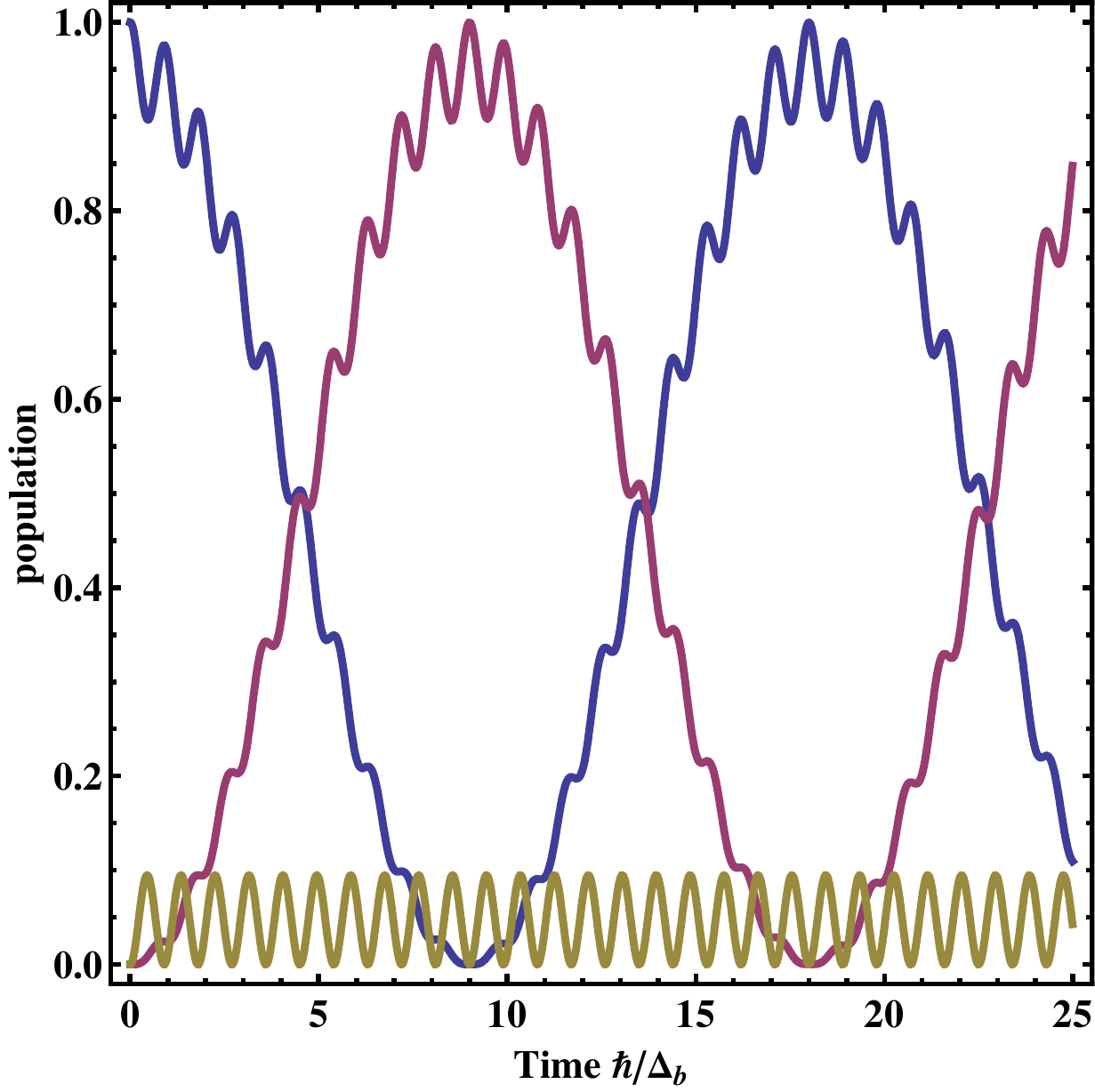}}
\caption{Time evolution of the
  population of super-plaquettes in  $N_{42}(t)$ (blue),  $N_{24}(t)$ (purple) and  $N_{33}(t)$  (yellow) configurations respectively. The fast frequency  component is determined by  $\Delta_b$ and the slow envelope by $J'^2 g^2/\Delta_b$. Here $U/J=2.6$ and $J'/J=0.01$.
}
\label{trace}
\end{figure}

Under coherent
quantum evolution, this system will exhibit Rabi oscillations between
$|4,2\rangle$ and $|2,4\rangle$ states with the $|3,3\rangle$ populated
virtually (as seen in Fig. \ref{trace}). The frequency of oscillation of the
envelope is given by $4g^2J'^2/\Delta_b$ and can be used to measure the binding
energy of the hole pairs. These ideas are similar to the ones used to measure
the super-exchange energy in an array of double wells\cite{rey}. We next
describe a method to measure the number of super-plaquettes
in $|4,2\rangle$, $|2,4\rangle$ and $|3,3\rangle$ states using
band-mapping techniques.

The distribution of super-plaquettes in different states can be
obtained from the number distribution of fermions on the left and right
plaquettes separately. To obtain the number distribution, after quenching
$J'$, each
plaquette is first converted to a single site by removing
the short lattice potential. By counting the number of the atoms in each
Brillouin zone after a band-mapping sequence on release from the lattice,
the average occupation of all bands is determined. In a second step,
the sequence is repeated, but the vibrational level of each atom on the
right plaquette is increased by one after the conversion to a single well.
This can be achieved in analogy to the scheme demonstrated in \cite{Porto2},
 by splitting the combined well and recombining in the presence of an energy
 bias.The bias is created using a phase shift on the 4-lambda lattice in
 such a way that only one of the plaquettes is affected.
 By comparing the
 resulting band occupations with those of step 1, the average filling of each
band on each side is reconstructed. Both of these steps can be repeated with
 an additional Feshbach resonance sweep which converts all atom pairs to
molecules. The remaining single atoms allow the separation of the fractions
of even and the odd number states in each band of each side, which completes
 the determination of the plaquette atom number statistics.

From the obtained histogram of atom number distributions for the left and
right plaquette at time $t$, $N_{42}(t)$ , $N_{24}(t)$ and $N_{33}(t)$ can
be determined\cite{hist_ftnt} to measure the hole-pair binding energy.

\section{Phase Diagram on Checkerboard Lattice}

\begin{figure*}
\centering
\subfigure[]{\includegraphics[width=2.5in]{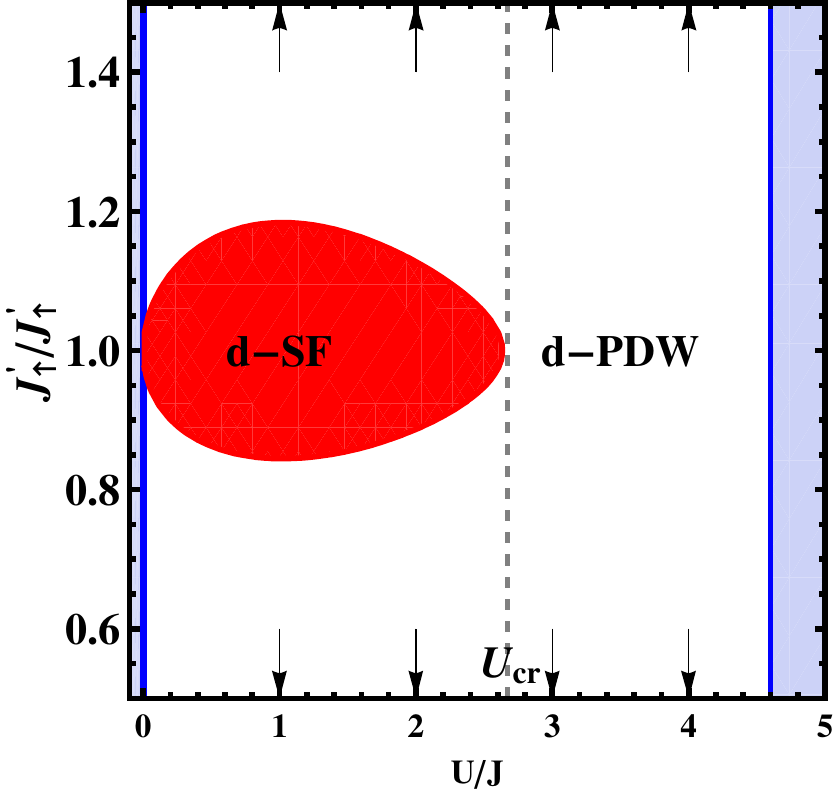}}\quad
\subfigure[]{\includegraphics[width=3.7in,height=2.3 in]{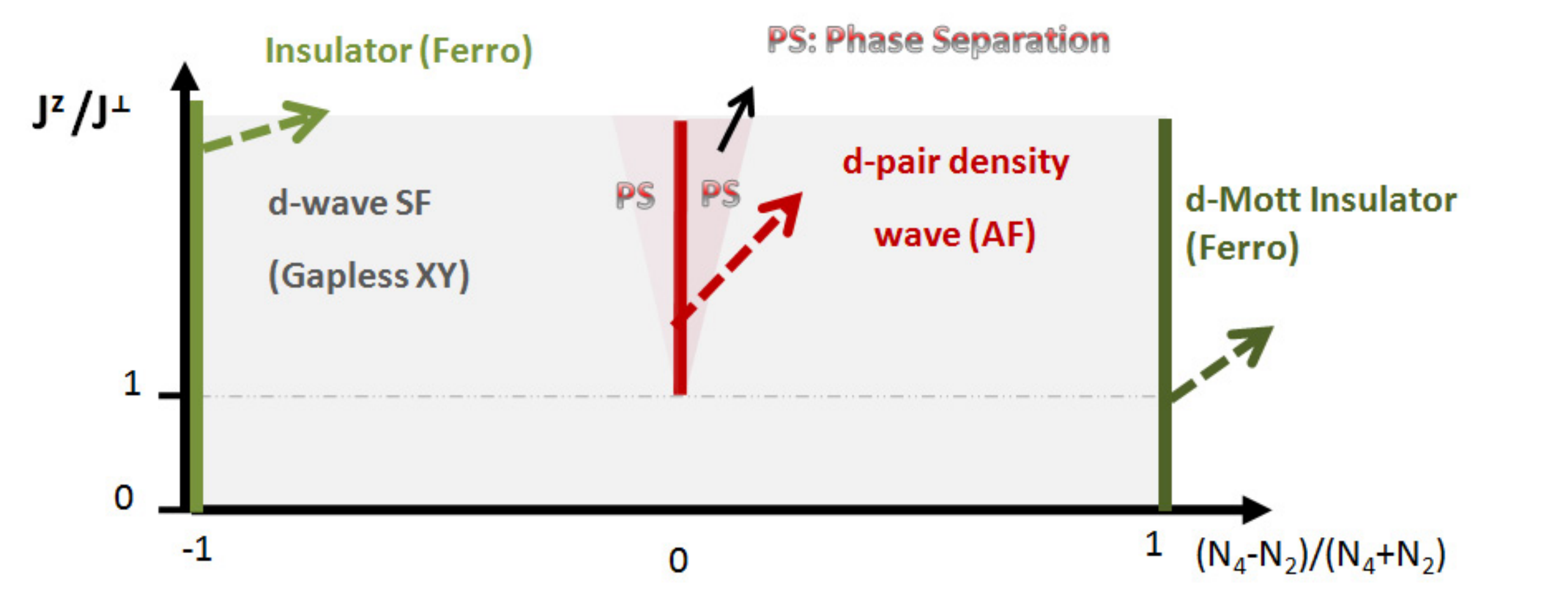}}
\caption{  a) Zero temperature phase diagram at $3/8$ fermionic filling
($N_4=N_2=N_{total}/2$). At this filling the system exhibits a second order
phase transition from a d-wave Superfluid, d-SF, (magnetic  ordered state
in the XY plane for the effective spins) to a d-pair density wave state,
d-pdw, (antiferromagnetic ordered phase for the effective spins ) when the
axial, $J^z$, and the  transverse $J^{\perp}$, coupling constants that appear
in the XXZ Hamiltonian become equal. As both $J^{\perp}$  and $J^z$ depend on
the  inter-plaquette tunnelings $J'_{\uparrow,\downarrow}$ and the ratio $U/J$,
the critical point  can be controlled by tuning these microscopic parameters.
b) Zero temperature phase diagram of the effective XXZ Hamiltonian as a
function of the  filling factor: Away from $3/8$ fermionic filling
($N_4=N_2$), the d-wave superfluid phase is energetically favorable except
at the fermionic filling factors $1/4$ (or $N_2=N_{total})$ and  $1/2$ or
($ N_4=N_{total})$ where the system becomes an insulator. There is an
additional first order phase transition from a d-pdw to a d-Sf as the
filling factor is varied away $N_4=N_2$ and this induces phase separation
in a small range of filling factors around $N_4=N_2$).} \label{para}
\end{figure*}

Now we study the more general case of a checkerboard array of plaquettes such as the one shown in Fig.\ref{superp}c.
We will restrict our analysis to the regime $|\Delta_b| \gg  g J_\sigma^{'}$  and $\Delta_b>0 $ where we can treat  the states  $|2\rangle$  and
  $|4\rangle$  as the low energy modes and adiabatically eliminate high energy states via second order perturbation theory.  This procedure yields a more general effective XXZ Hamiltonian given by

\begin{eqnarray}
\displaystyle H_{eff}=\sum_{<{\bf{R,R'}}>}\left[-J^{\perp}(\sigma_{\bf{R}}^x\sigma_{\bf{R'}}^x+\sigma_{\bf{R}}^y\sigma_{\bf{R'}}^y)+J^{z}
\sigma_{\bf{R}}^z\sigma_{\bf{R'}}^z\right] \label{xxz}
\end{eqnarray}
where $J^{\perp}=\frac{ g^2 J_\uparrow'J_\downarrow'
}{\Delta_b}$, $ J^{z} =(J_\uparrow'^2+ J_\downarrow'^2 )\frac{ g_z^2}{2\Delta_b} $. Here we have neglected terms proportional to  $\sum_{\bf{R}} \sigma_{\bf{R}}^z$
since it is a
conserved quantity in these systems. The coupling matrix element $g_z$ has a
complicated dependence on $U/J$ as shown in Fig.\ref{bindi}. For details
about its derivation see methods.

The zero temperature phase diagram of the XXZ Hamiltonian is known and
consequently can be used to infer the phase diagram of the corresponding
fermionic system \cite{Hebert,Batrouni,Kivelson}. At $3/8$ fermionic
filling, i.e. $N_4=N_2$, there is a second order phase transition as the
$J^z/J^{\perp} $ ratio is varied: While for
$J^z/J^{\perp} <1$ the ground state corresponds to  a gapless  d-wave
superfluid ( a magnetically ordered phase in the XY-plane for the
effective spins), for $J^z/J^{\perp}  >1$  it becomes a gapped Cooper pair
density wave state, d-pdw (an antiferromagnetic order phase for the
effective spins).
The pdw state is not the usual particle-hole charge density wave state, but
can be viewed as a crystal of d-wave Cooper pairs \cite{Cooperxtal}.
The point  $J^z=J^{\perp}$ is the critical point.  In Fig.~\ref{para}a, we
show the phase diagram  as a function of $J'_\uparrow/J'_\downarrow$ where the
tendency of anisotropic tunneling to stabilize the d-pdw phase can be observed.
For the spin independent case $ J'_\uparrow= J'_\downarrow$, our phase diagram
is in agreement with the one obtained in Ref.\cite{Kivelson}, exhibiting a
critical point at  $U_{cr}\sim 2.7 J$. Notice that since our analysis is
based on the assumption that  $g J'\ll \Delta_b$, the parameter regime where
 it is  applicable considerably reduces  as one approaches the points $U=0$
and $U/J\sim 4.6$ where $\Delta_b$ vanishes.

Away from  $3/8$ fermionic  filling, the  phase diagram  is almost
insensitive to the $J'_\uparrow/J'_\downarrow$ ratio. There is a first order
phase transition from the d-pdw phase to the d-wave superfluid as
 the chemical potential is varied away from $3/8$ filling. Due to the first
order character of the transition, the d-pdw phase is surrounded by a small
region where one observes phase coexistence \cite{Hebert,Kivelson}. Except
this region and the state at special filling factors: $1/4$, $3/4$, $1/2$
and $1$ when the system turns into an insulator,
the low energy phase is always a gapless d-wave superfluid as shown
in Fig.~\ref{para}b.

\begin{figure*}[htp]
\begin{center}
\leavevmode {\includegraphics[width=6 in]{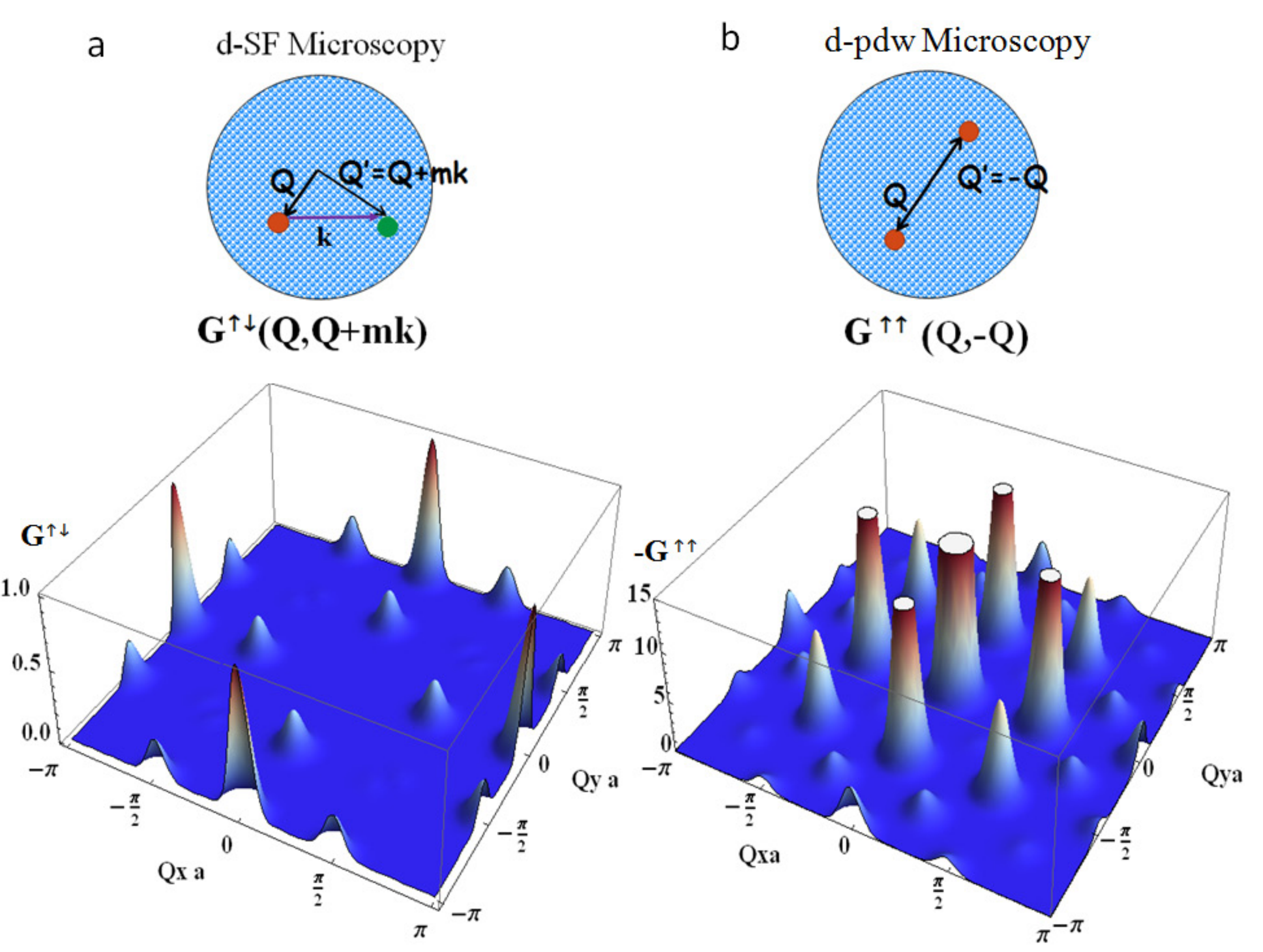}}
\end{center}
\caption{The left  panels display the unequal spin  noise correlations, $ G^{\uparrow\downarrow}(\textbf{Q} ,\textbf{Q'}=\textbf{Q}+m\textbf{k} )$, where $\textbf{k}=2\pi/a \hat{x}$ is  the reciprocal lattice vector of the  underlying lattice with spacing $a$, assuming   a $d-$ wave  superfluid. The right panel shows  equal spin noise  correlations (with opposite sign)  $ -G^{\uparrow\uparrow}(\textbf{Q} ,\textbf{Q'}=-\textbf{Q})$   for a d-pdw state. Here we have subtracted  the local density-density correlations within a plaquette (see methods).   $ G^{\uparrow\downarrow}(\textbf{Q} ,\textbf{Q'} )$  exhibits  interference peaks  at $\textbf{Q}+\textbf{Q'}= \textbf{K} m$ in the d-wave superfluid regime with $\textbf{K}=\textbf{k}/2$  is  the reciprocal lattice vector of the plaquette array. The peaks are modulated by an overall envelope  which  probes the $d$ wave symmetry.
This modulation  causes the disappearance of the peaks   along the nodal lines $Q_x=\pm Q_y$. On the contrary $- G^{\uparrow\uparrow}(\textbf{Q} ,-\textbf{Q} )$   exhibits  interference peaks  at $\textbf{Q}-\textbf{Q'}= \textbf{K} (2m+1)/2$ in the d-pdw state. Due to the large kinetic energy of the fermions within a plaquette  these peaks however are much weaker than the anti-bunching peaks, which always appear at  $\textbf{Q} +\textbf{Q'}=n \textbf{K}$.
 }\label{noised}
\end{figure*}

To explore the different quantum phases, we propose to first
prepare an initial array of isolated plaquettes, with the desired filling
factor (see methods), and then slowly increase the inter-plaquette coupling
taking advantage of the experimental ability to tune the lattice geometry
and the ratio $J^z/J^{\perp}$ by using spin dependent control. The effective
XXZ Hamiltonian allows us to estimate the entropy of the ordered state
(near $T_c$) and hence the  $T/T_F$ of the initial trapped
Fermi gas without the optical lattice required for observing these phases
(assuming all later modifications are adiabatic). This gives
an estimate of $T/T_f\sim 0.01$ to observe the d-wave superfluid and
$T/T_f\sim 0.03$ to observe the charge ordered state. These values should
be achievable with current technology.

\section{Detection of the quantum phases via Noise correlations}
Noise correlations \cite{Altman} can be used to
detect the different quantum phases. The atomic noise in an expanding cloud
is related to the following four point functions at the time of the release:
\begin{eqnarray}
G^{\sigma \sigma'}_{\textbf{Q,Q'}}\propto \langle \hat{n}_{\textbf{Q}\sigma}
\hat{n}_{\textbf{Q}'\sigma'}\rangle
- \langle  \hat{n}_{\textbf{Q}\sigma}\rangle \langle
\hat{n}_{\textbf{Q}'\sigma'}\rangle
\end{eqnarray}
with $\hat{n}_{\textbf{Q}\sigma}\propto \sum_{l,s}e^{ i \textbf{Q} \cdot (
\textbf{L}_{ls})}\langle  c_{l\sigma}^{\dagger} c_{s\sigma}\rangle$, being the
quasi-momentum distribution and $\textbf{L}_{ls}$ a vector connecting the
lattice sites $s$ and $l$. $G^{\sigma \sigma'}(\textbf{Q}',\textbf{Q})$ has  the required ingredients to distinguish the two phases. On one hand, $G^{\uparrow \downarrow}(\textbf{Q},\textbf{Q}')$ contains  terms proportional to $|\langle\hat{\Delta}\rangle|^2$ (see Eq.\ref{dwave}), and consequently a d-wave superfluid  with  $\langle \hat{\Delta}\rangle\neq 0$ must exhibit interference fringes
at $\textbf{Q} + \textbf{Q}'= \textbf{K} n$,   $\textbf{K}$  being the reciprocal lattice vector of the plaquette array, which is half of the reciprocal lattice vector of the  underlying lattice.  The $d$ wave nature of the state will be signaled by a modulation of the peaks with an  a overall envelope with the characteristic d-wave nodal planes along  $Q_x=\pm Q_y$ and   $Q'_x=\pm Q'_y$ as shown in Fig.\ref{noised}.
Information about the  density order is given by  $G^{\uparrow\uparrow}$  which will  show  sharp dips at
$\textbf{Q} - \textbf{Q}'=(2n+1)\textbf{K}/2$   in the presence of
a d-pdw phase and a  flat profile in the superfluid phase.  In contrast to the unequal spin
correlations,   $G^{\uparrow\uparrow}$ will  not have the inherent  $d-wave$ symmetry. Equal spin correlations will also always exhibit
dips at   $\delta(\textbf{Q} -\textbf{Q} '- n\textbf{K})$ which  reflect the characteristic anti-bunching of fermions \cite{Rom}. Unfortunately, the amplitude of these dips can be   about 30 times stronger than the   dips  at $(2n+1)\textbf{K}/2$ inherent from the charge density order (see Fig. \ref{noised}b). This caveat  can be avoided  by quenching the  kinetic energy of the atoms in the plaquettes before the release  by slowly merging  each plaquette into a single well. For details about the noise interference pattern we refer the reader to the method section.

\section{Conclusions}

In summary we have described a technique to prepare and  detect
 d-wave superfluidity   in ultra-cold  fermionic systems loaded
in   optical  super-lattices. Even though our theoretical predictions are  restricted to  a regime where
perturbative expansions are valid, the ideas presented in this letter    might open a window to explore in the laboratory more complex  regimes  which do not allow  for a clear theoretical description.
We  point  out that so far we have ignored the presence of any  any inhomogeneous confining potential in the checkerboard lattice case.
As the binding energy is a very sensitive quantity to any bias within a plaquette, its presence can certainly prevent the observation of the predicted phases. Consequently in this type of experiments it would be ideal to use a   "box" type trapping  potential such as the one proposed in Ref.\cite{Raizen}

This work was supported by ITAMP, NSF (Career Program), Harvard-MIT CUA, AFOSR, Swiss
NF, the Sloan Foundation, and the David and Lucille Packard
Foundation.

\section*{Methods}
\subsection*{ Beyond the Effective Hamiltonian in a plaquette}

The validity of the effective Hamiltonian, Eq.( \ref{tpl}) can be checked in the super-plaquette system
  by going one step beyond and deriving a more general effective Hamiltonian within  the Hilbert subspace spanned  by all the direct products of the low energy eigenstates of the
isolated plaquette i.e $|4,2\rangle$,    $|2,4\rangle$ and eight possible
$|3^{(\sigma,\tau)},3^{(\bar{\sigma},\bar{\tau})}\rangle$ configurations. Even though in total there are $10$
states  only 4 of them are coupled by $J'_\sigma$. These are
$|t\rangle$,
$|+\rangle=\sum_{\sigma\tau}(-1)^{\sigma_z-1}|3^{(\sigma,\tau)},3^{(\bar{\sigma},\bar{\tau})}\rangle/2$,
 $|s\rangle$ and
$|-\rangle=\sum_{\sigma\tau}|3^{(\sigma,\tau)},3^{(\bar{\sigma},\bar{\tau})}\rangle/2$. In this basis the
generalized effective Hamiltonian  reduces to

\begin{equation}
H=\left(
    \begin{array}{cccc}
      0 & - g \bar{J'} & 0  & 0 \\
     -g \bar{J'}  & \Delta_b & 0 & 0 \\
     0  & 0 &  0 &-g \delta J'  \\
      0& 0 & - g  \delta J' &\Delta_b
      \end{array}
  \right)\label{matrix}
\end{equation}
Here  $\delta J'=J^{'\uparrow}-J^{'\uparrow}$ and
$\bar{J'}=J^{'\uparrow}+J^{'\downarrow}$. The Hamiltonian is  block diagonal
so the eigenstates  are independent  linear combinations of
$\{|t\rangle, |+\rangle\}$  and $\{|s\rangle,|-\rangle\}$ respectively and
have energies:
\begin{eqnarray}
\nonumber \hbar\omega_{1,3}=\frac{\Delta_b\mp \sqrt{\Delta_b^2 + 4g^2 \bar{J'}^2} }{2} ~~~\\
\hbar\omega_{2,4}=\frac{\Delta_b\mp \sqrt{\Delta_b^2 +4 g^2 \delta J'^2 }}{2}.
\end{eqnarray}
For  $g J'_\sigma \ll |\Delta_b|$ the eigenstates are grouped in two doublets
with common  intra-doublet splitting $4g^2 J^{'\uparrow}J^{'\downarrow}/\Delta_b$
and doublet splitting $\sim\Delta_b$. The sign of  $\Delta_b$ determines
which one of the two doublets is higher in energy. Since first order
tunneling transfers amplitudes across the two doublets, if
$g J'_\sigma \ll |\Delta_b|$, these processes are energetically costly and
only second order processes within the doublets are relevant. This assumption
yields the effective spin Hamiltonian, Eq.(\ref{tpl}). On the contrary for
$g J'_\sigma \sim  |\Delta_b|$ there is no clear energy separation between the
doublets and the ground state is a mixture of the various independent
plaquette states.

The corrections from the effective Hamiltonian can be measured by probing
the coherent dynamical evolution proposed in the main text.
In the weak coupling regime $N_{24}(t)$ and $ N_{42}(t)$ will exhibit nice
sinusoidal oscillations with a frequency proportional to $ g^2J'^2 /\Delta_b$.
However for $\Delta_b \sim  J'$ a more complex dynamics will occur.
More quantitatively, using Eq.(\ref{matrix}) one gets that
\begin{eqnarray}
N_{24,42}(t)&=&\frac{1}{2} -4 g^2 J'^2 \frac{\sin^2[\omega_{13}t/2]}{\hbar^2\omega_{13}^2}\notag\\
& &\pm\frac{\omega_1 \cos(\omega_3 t)-\omega_3 \cos(\omega_1 t)}{2\omega_{13}} \label{anal}\notag\\
N_{33}(t)&=&8 g^2 J'^2 \frac{\sin^2[\omega_{13}t/2]}{\hbar^2\omega_{13}^2}
\end{eqnarray}
Here $\omega_{13}=\omega_{1}-\omega_{3}$. In the limit  $\Delta_b\gg J'$,   $\omega_3$ dominates over  $\omega_{1}$, the $|3,3\rangle$ states are only virtually populated and the above expressions reduce to
\begin{eqnarray}
N_{24,42}(t)=\frac{1}{2}\left(1 \pm\cos(\omega_1 t)\right) \quad N_{33}(t)\thickapprox 0
\end{eqnarray}

\subsection*{Derivation of the Effective Hamiltonian parameters: Multi-plaquette case}
The microscopic parameters of the many-plaquette Hamiltonian are more
complicated that the ones in Eq.(\ref{tpl})
since in addition to the virtual couplings between
$|4_{R},2_{R+1}\rangle\Leftrightarrow|2_{R},4_{R+1}\rangle$
through tunneling to $|3_{R},3_{R+1}\rangle$ (which are inversely proportional
to $\Delta_b$), when many plaquettes are coupled, there are additional
processes that one must consider: We have to include the
couplings between $|4_{R},4_{R+1}\rangle\Leftrightarrow|4_{R},4_{R+1}\rangle$
mediated by $|5_{R},3_{R+1}\rangle$ states and between
$|2_{R},2_{R+1}\rangle\Leftrightarrow |2_{R},2_{R+1}\rangle$
mediated by $ |3_{R},1_{R+1}\rangle$ (which are  inversely proportional to
$g_4^2 /\Delta_{44}$ and $g_2^2/ \Delta_{44}$ where
$\Delta_{44}=E_5+E_3-2E_4$, $\Delta_{22}= E_3+E_1-2E_2$ and $g_{4,2}$ are
corresponding coupling matrix elements). Since the energy gaps $\Delta_{44}$
and $\Delta_{22}$ are about one order of magnitude larger than $\Delta_b$,
to be consistent, one has in addition to sum over all possible  higher energy
virtually populated states (leading to additional corrections we denote
by $\delta$). All these processes give rise to an effective coupling
constant along the axial direction given by:
$ g_{z}^2 =g^2-\frac{g_2^2 \Delta_b }{2\Delta_{22}}  - \frac{ g_4^2 \Delta_{b}}{2\Delta_{44}}+ \delta $

\subsection*{Initial preparation of the plaquettes with $N_2=2N_4$}

 The  preparation starts by loading a unit filled band insulator  of
fermionic atoms in a rectangular lattice with lattice period
$3/2 \lambda$ and $\lambda$ along the $x$ and $y$ direction respectively.
After this, the $\lambda$ lattice is ramped up  along x simultaneously with a weaker $3\lambda$ lattice. While the former creates an additional
well in between two initially populated wells, the latter provides an energy
offset to guarantee that the new wells remain unpopulated. Next, one has to
slowly turn off the $3/2 \lambda$ lattice while turning on a $ 2\lambda$
lattice. At the end of this procedure  one has  generated a distribution of
double wells along x with the appropriated bias such that the ground state
population has $242$ pattern. After increasing the lattice depth of the
$2\lambda$ lattice until all wells are made independent one can suddenly
turn off the $ 3\lambda$ lattice which provided the bias without causing any
excitation. The overall scheme creates an array of independent plaquettes
loaded in the ground state with alternating filling factors of $2,4, 2$ along
the x direction.

\subsection*{Noise Correlations}

Here we provide explicit expressions for the noise correlations shown in
Fig.\ref{noised}.
These expressions can be derived using the fact that the states to probe are linear superpositions of states   populated with only 4 and 2 fermions in each plaquette, and therefore that they can be mapped  to an  effective spin $1/2$ system.
Using this simplification, the noise correlations can be written as
 \begin{eqnarray}
G^{\uparrow\downarrow}(\textbf{Q} ,\textbf{Q} ') &\propto& A(\textbf{Q} ,\textbf{Q} ') +\bf{S}^{+-}(\textbf{Q} +\textbf{Q} ') \mathcal{D}(\textbf{Q} ,\textbf{Q} ')\\
G^{\uparrow\uparrow}(\textbf{Q} ,\textbf{Q} ') &\propto& A(\textbf{Q} ,\textbf{Q} ')-F^+(\textbf{Q} ,\textbf{Q} ')\delta(\textbf{Q} -\textbf{Q} '- m \textbf{K}) \notag \\ &&-S^{zz}(\textbf{Q}
-\textbf{Q} ') F^-(\textbf{Q} ,\textbf{Q} ')
\end{eqnarray}

The term $A$  corresponds to trivial diagonal density-density correlations
within a plaquette and ideally one could subtract this term from the measured
noise pattern.
$S^{+-}(\textbf{Q}+ \textbf{Q}')=\sum_{\textbf{R} \neq\textbf{R'}} e^{ i (\textbf{Q} + \textbf{Q}' ) \cdot ( \textbf{R}-\textbf{R'})}
\langle \sigma_{\textbf{R}}^+ \sigma_{\textbf{R}'}^-\rangle$
is the  analog of the quasi-momentum distribution
when one maps the spins to hard-core bosons and consequently probes the long
range order. If the system is a $d-wave$ superfluid
$S^{+-}(\textbf{Q}+ \textbf{Q}')$ will exhibit sharp interference peaks
at $\textbf{Q} + \textbf{Q}'= \textbf{K} n$, $\textbf{K}$ being the
reciprocal lattice vector of the plaquette array, which is half of the
reciprocal lattice vector of the  underlying lattice. On the contrary
$S^{+-}(\textbf{Q}+ \textbf{Q}')$ will have a flat profile in the case of a
d-cdw state. The term
$\mathcal{D}(\textbf{Q} ,\textbf{Q} ')=\left |\sum_{r  r'} e^{ i
(\textbf{Q} \cdot \textbf{l}_{r}+\textbf{Q}'\textbf{l}_{r'})} \langle2|c^{\downarrow \dagger}_{{r}}
c^{\uparrow\dagger}_{{r'}}|4\rangle\right |^2 $
probes non-diagonal local plaquette correlations
($r,r'$ label sites within the same plaquette) and
gives information about  the $d-$ wave nature of the superfluid.
 $\mathcal{D}(\textbf{Q} ,\textbf{Q} ')$
modulates the $\bf{S}^{+-}(\textbf{Q} +\textbf{Q} ')$ peak structure inducing
the characteristic d-wave nodal lines along  $Q_x=\pm Q_y$ and
$Q'_x=\pm Q'_y$, where the signal vanishes due to the d-wave symmetry of the
Cooper pairs (see Fig.\ref{noised}).

Information about the  density order is given by  $G^{\uparrow\uparrow}$.
$S^{zz}(\textbf{Q}- \textbf{Q}')=\sum_{\textbf{R} \neq\textbf{R'}} e^{ i (\textbf{Q} - \textbf{Q}' ) \cdot (
\textbf{R}-\textbf{R'})} \langle \sigma_{\textbf{R}}^z \sigma_{\textbf{R}'}^z\rangle$  shows  sharp peaks at
$\textbf{Q} = \textbf{Q}'+(2n+1)\textbf{K}/2$   in the presence of
a d-cdw phase and a  flat profile in the superfluid phase.  In contrast to the unequal spin
correlations the terms $F^{\pm}(\textbf{Q} ,\textbf{Q} ')$ do not have the inherent  $d-wave$ symmetry. They are just geometric factors
that depend on the kinetic energy  within  the isolated plaquettes : $F^{\pm}(\textbf{Q}
,\textbf{Q} ')=|\alpha_4(\textbf{Q} ,\textbf{Q} ')\pm-\alpha_2(\textbf{Q} ,\textbf{Q} ')|^2/4 $  with
$\alpha_n(\textbf{Q} ,\textbf{Q} ')=\sum_{r  r'} e^{ i (\textbf{Q} \cdot \textbf{l}_{r}-\textbf{Q}' \cdot\textbf{l}_{r'})}
\langle n|c^{\sigma \dagger}_{{r}} c^{\sigma}_{{r'}}|n\rangle$, and $n=2,4$.
The delta function   $\delta(\textbf{Q} -\textbf{Q} '- m \textbf{K})$ reflects the characteristic anti-bunching of fermions.

\begin{thebibliography}{99}
\bibitem{hist_ftnt} This assumes a good initial preparation, i.e. only the
$|4,2\rangle$, $|2,4\rangle$ and $|3,3\rangle$ configurations are populated.
In realistic situation defects might be present and additional measurements
will be needed for a full reconstruction  of left/right histograms.
\end{thebibliography}
\end{document}